\documentclass[preprint,aps,showpacs,pre]{revtex4}
\usepackage[dvips]{graphicx}
\usepackage{color}

\begin{document}

\title{Adsorption-desorption phenomena  and diffusion of neutral particles in the hyperbolic regime}

\author{A. Sapora$^{1,2}$, M. Codegone $^{1,3}$, G. Barbero$^{1,4}$, and L. R. Evangelista$^5$}
\affiliation{$^1$ Turin Polytechnical University in Tashkent,
 17, Niyazov Str. Sobir Rakhimov district Tashkent, 100095 Uzbekistan.\\
$^2$ Department of Structural, Building and Geotechnical Engineering\\
Politecnico di Torino Corso Duca degli Abruzzi 24
10129 Torino - Italy \\
$^3$ Department of Mathematical Sciences,
Politecnico di Torino, Corso Duca degli Abruzzi 24, 10129 Torino, Italy.\\
$^4$Department of Applied Science and Technology,
Politecnico di Torino, Corso Duca degli Abruzzi 24, 10129 Torino, Italy.\\
$^5$ Departamento de F\'{i}sica, Universidade Estadual de Maring\'a,
Avenida Colombo 5790, 87020-900 Maring\'a, Paran\'a, Brazil.}
\begin{abstract}
The adsorption phenomenon of neutral particles from the limiting surfaces of the sample, in the Langmuir's approximation, is investigated.
The diffusion equation regulating the redistribution of particles in the bulk is assumed of hyperbolic type, in order to take into account
the finite velocity of propagation of the density variations. We show that in this framework the condition on the conservation of the
number of particles gives rise to a non local boundary condition. We solve the partial differential equation relevant to the diffusion of
particles by means of the separation of variables, and  present how it is possible to obtain approximated eigenvalues contributing to the solution.
The same problem is faced numerically by a finite difference algoritm. The time dependence of the surface density of adsorbed particles is deduced by
means of the kinetic equation at the interface. The predicted non-monotonic behavior of the surface density versus the time is in
agreement with experimental observations reported in literature, and it is related to the finite velocity  of propagation of the density variations.
\end{abstract}
\date{\today}
\maketitle

\section{Introduction}
The standard analysis of adsorption phenomenon is made by means of the diffusion equation, in the framework of the Langmuir
approximation~\cite{Libro}. In this scenario, to
dynamically describe the adsorption-desorption process in confined systems, the diffusion equation has to be
solved  to obtain the bulk density of
particles,
 and a kinetic balance equation has to be imposed at the interface to determine the time dependence of
the particles surface density~\cite{EBPRE,Libro}.
For some systems, the temporal behavior of the surface density of adsorbed particles may be non-monotonic, presenting some kind of
oscillatory behavior, and tending to a saturation value only for large enough time~\cite{Cosgrove}. This may happen because, in the complex
adsorption phenomenon occurring at a solid surface,
the collision of a molecule can be represented by at least three different
processes. The simplest one is an elastic scattering that occurs when there is
no loss of translational energy during the collision of the molecule at the interface. Frequently, however,
if the molecule is in a weakly bound state,  then the thermal motion of the surface atoms can
cause the molecule to undergo desorption. Finally, when the molecule
collides with the surface, it may lose energy and is converted into
a state where it remains on the surface for a reasonable time, i.e., it sticks~\cite{Masel}.
Thus, for a more general process it is reasonable to assume that the actual
position of the molecule on the surface
has some kind of  ``memory'' of its incoming state, eventually modifying
the adsorption-desorption rates. In order
to incorporate memory effects in the adsorption phenomena, one way is to
propose a modified kinetic equation in which the suitable choice
for a temporal kernel in the desorption rate can account for the
relative importance of physisorption or of chemisorption, according
to the time scale governing the adsorption phenomena~\cite{WEPRE,CPL}.
This approach focuses on the surface behavior.
Another way to face the problem is to consider that the density becomes a non-Markovian process and the diffusion equation is given by a
persistent-random-walk process~\cite{Colin}. To account for this new effect, the current density of diffusing particles is assumed to satisfy the non-Fickian
Maxwell-Cattaneo relationship. Indeed,
in 1948 Cattaneo has proposed a new constitutive equation that contains a new term and changes the Fourier equation into a hyperbolic one,
whose solution propagates at finite velocity~\cite{Cattaneo,Preziosi}.
In this modified scenario, the behavior of the bulk density of particles is governed by the Cattaneo's equation, while the usual kinetic
equation governs the adsorption-desorption rates at the interface (Langmuir's approximation). This implies that the diffusional process
will not be described in the parabolic regime, as is usually done in the classical  Fourier approach for the heat propagation, but will
be considered in the hyperbolic regime~\cite{Wang}. In this framework, due to the finite velocity propagation of the solution in a confined sample,
one expects that a non-monotonic behavior for the density of particles at the surface may be found for suitable values of the parameters
entering the model.  Recently, a comparative analysis of the
predictions of the diffusive models in the hyperbolic and parabolic regimes has been carried out theoretically~\cite{Threetenors}.  By employing two types of  initial gaussian-type distribution of the diffusing particles, namely, one centered
around the symmetry surface in the middle of the sample and another one localized close to the limiting surfaces, it was shown that the evolution towards to
the equilibrium distribution is not monotonic.

An alternative approach that considers the diffusion processes in a more general framework may be represented by the use of fractional operators, i.e. derivatives with non-integer orders~\cite{Carpinteri}. In this case, the fractional exponent of the time derivative is assumed to lay between 1 and 2, representing the two limit situations of a parabolic equation and a hyperbolic equation. In describing anomalous transport in the framework of the fractional dynamics~\cite{Restaurant}, even generalized Cattaneo equation may be used to approach anomalous transport processes~\cite{Compte}. In general, anomalous behavior is usually related to the non-Markovian~\cite{Yednak} characteristics of the systems such as memory effects, fractality, and interactions. This rich class of phenomena may be conveniently faced by the methods recently proposed to investigate anomalous diffusion employing the techniques of  the fractional calculus and its applications in physics~\cite{Metzler1,Metzler2,Metzler3,Hilfer1,Nonnemacher1,Hilfer2,Schot,dePaula} .  Thus, the physical scenarios represented by Cattaneo-like approaches and fractional calculus may be connected to explore non usual diffusive behavior.  In this optic, the present analysis may reveal to be useful also to generalize the results recently found investigating the diffusion process in nonlocal solid
mechanics~\cite{Carsap,Sapora}.

The organization of the paper and the main achievements are as follows.   In Sec.~\ref{position}, the statement of the mathematical
problem is done in terms of the usual diffusion equation and the kinetic equation at the interface is introduced. The condition stating the
conservation of the number of particles
is also imposed and a set of fundamental equations governing the parabolic regime is established. In Sec.~\ref{Generalization}, the
problem is reformulated in terms of a hyperbolic diffusion equation to account for the finite velocity of propagation of the density
variations. There it is shown that the condition on the conservation of the number of particles leads to a non-local boundary
condition. In Sec.~\ref{Eigenvalues}, we formally solve the partial differential equation relevant to the diffusion of particles by means of
the separation of variables, and show how it is possible to obtain approximated values for the eigenvalues for the considered problem. In
Sec.~\ref{Initial},  analytical results for the bulk and surface densities are obtained by means of an orthogonalization process that is
also discussed in some detail. Since the same problem is numerically solved (Sec.~\ref{Numerical}), we  dedicate Sec.~\ref{Results} to present some relevant
results obtained by the two strategies employed in the paper. The section ends with some concluding remarks on the predictions of the
hyperbolic model and its relevance in experimental contexts.

\section{Position of the problem}
\label{position}

In the following, we will consider the simple case of a confined sample in the shape of a slab of thickness $d$,  limited by two
identical, adsorbing,  surfaces. In this case the problem is one-dimensional. We use a cartesian reference frame having the $z$-axis
perpendicular to the limiting surfaces, placed at $z=\pm d/2$. We assume that  at $t=0$ the distribution of particles across the sample is
constant, characterized by a bulk density $n_0$, and the adsorbing surfaces are not working. For $t=0$,  the surfaces start to adsorb the
particles, and the bulk density of particles changes with $t$, until  reaching a new equilibrium distribution, $n_{\rm eq}$. We assume
that the adsorption phenomenon is related to a localized (short range) energy due to the surfaces. In this situation, the bulk density of
equilibrium is homogeneous across the sample, and the adsorption phenomenon is related to the surface density of adsorbed particles,
$\sigma$. Hence,  in the case we are considering $n(z,t)$ passes from $n_0$ to $n_{\rm eq}$ and $\sigma$ from $0$ to $\sigma_{\rm eq}$.
In this case the fundamental equation of the problem is

\begin{equation}
\label{1}\frac{\partial n}{\partial t}=D\,\frac{\partial ^2 n}{\partial z^2},
\end{equation}
for the bulk, that has to be solved with the kinetic equation at the surfaces
\begin{equation}
\label{2}\frac{d \sigma}{dt}=k_a n_s(t)-\frac{1}{\tau_a}\,\sigma(t),
\end{equation}
where $n_s(t)=n(\pm d/2,t)$ is the density of particles just in front of the limiting surfaces, and $k_a$ and $\tau_a$ are two phenomenological
parameters related to the adsorption phenomenon, in the approximation of Langmuir.
Since the particles cannot leave the sample we have also the condition

\begin{equation}
\label{3-1}\int_{-d/2}^{d/2} n(z,t) dz+2 \sigma(t)=n_0 d,
\end{equation}
in our symmetric problem, where $\sigma(t)$ is the surface density of adsorbed particle on the surfaces at $z=\pm d/2$.
In the simple case where the bulk differential equation is the diffusion equation, (\ref{1}), Eq.~(\ref{3-1}) is equivalent to
\begin{equation}
\label{4-4-1}-D\left(\frac{\partial n}{\partial z}\right)=\frac{d \sigma}{dt},
\end{equation}
at $z=\pm d/2$. Note that with our hypotheses $n(z,t)=n(-z,t)$, and the two boundary conditions (\ref{4-4-1}) reduce to only one equation, which
is of local type. This means that  $(\partial n/\partial z)(\pm d/2,t)$ depends only on surface density $\sigma_{\pm}(t)$.

The solution of this problem predicts a monotonic increasing of $\sigma$ from 0 to $\sigma_{\rm eq}$~\cite{Libro}. On the contrary,
experimental observations indicate a non-monotonic increasing of $\sigma=\sigma(t)$\cite{Adamson,Cosgrove}.

\section{Hyperbolic Generalization}
\label{Generalization}
A theoretical attempt to interpret the experimental data has been done by modifying the kinetic equation at the limiting
surfaces~\cite{WEPRE,CPL,Yednak}. As stated above, according to this point of view, the non monotonic behavior of the surface density of adsorbed particles versus the time has a surface origin, and it is taken into account by a modification of the standard adsorption isotherm of Langmuir. Our aim is to show that a non monotonic behavior of $\sigma=\sigma(t)$ can have a bulk origin, related to the finite velocity of the bulk variation density.

As it is well known, Eq.~(\ref{1}) is based on the assumption that the transmission of bulk
density takes place with infinite velocity~\cite{Cussler}. For this reason we modify Eq.~(\ref{1}) in an equation of hyperbolic type as
follows \cite{Cattaneo}
\begin{equation}
\label{6}\tau_r\frac{\partial^2 n}{\partial t^2}+\frac{\partial n}{\partial t}=D\,\frac{\partial ^2 n}{\partial z^2},
\end{equation}
where $\tau_r$ is a characteristic time related to the medium in which the particles are dispersed, and $c=\sqrt{D/\tau_r}$ is the velocity of the bulk variations of density. The fundamental
equations of our problem are Eq.~(\ref{6}), that has to be solved with Eq.~(\ref{2}) describing the adsorption phenomenon, and Eq.~(\ref{3-1}),
stating the conservation of the number of particles. Note that in the present case, from Eq.~(\ref{3-1}), does not follow a boundary condition
of the type (\ref{4-4-1}) because the density of current is no longer given by $j=-D(\partial n/\partial z)$.  In other words, when the fundamental equation of the problem
is Eq.~(\ref{6}), Eq.~(\ref{3-1}) cannot be reduced to a local boundary condition.

We write Eqs.~(\ref{2}),~(\ref{3-1}), and~(\ref{6}) in dimensionless form. To this end we introduce the quantities
$N=n d$, $L=k_a\tau_a/d$, $A=\tau_a/\tau_D$, $B=\tau_r/\tau_D$,
and the reduced coordinates
${z^*}=z/d$, and ${t^*} =t/\tau_D$, where $\tau_D=d^2/D$ is the standard diffusion time. Note that $\ell=k_a \tau_a$ is an intrinsic length related to the adsorption. In terms of the reduced coordinates the equations of the problem are

\begin{equation}
\label{9}B\,\frac{\partial^2 N}{\partial {t^*} ^2}+\frac{\partial N}{\partial {t^*} }=\frac{\partial^2 N}{\partial {z^*}^2},
\end{equation}
for $-1/2 \leq {z^*} \leq 1/2$, with the kinetic equation
\begin{equation}
\label{10} A\,\frac{d \sigma}{d {t^*} }=L\, N_s(t^*)-\sigma(t^*),
\end{equation}
where $N_s(t^*)=N(\pm 1/2,t^*)$, and the equation stating the conservation of particles
\begin{equation}
\label{11}\int_{-1/2}^{1/2} N({z^*},{t^*} )\,d{z^*}+2 \sigma(t^*)=N_0,
\end{equation}
where $N_0=n_0 d$. For ${t^*} =0$ the surface density of particle is such that $\sigma(0)=0$. Consequently the initial time derivative of $\sigma$, as it follows from Eq.~(\ref{10}), is
\begin{equation}
\label{10-1}\left(\frac{d \sigma}{d{t^*} }\right)_{0}=\frac{L}{A}\,\,N_s(0),
\end{equation}
independent of $B$.

In the final state of equilibrium, for ${t^*} \to \infty$, $N\to N_{\rm eq}$ and $\sigma \to \sigma_{\rm eq}$. These quantities, according to Eqs.~(\ref{10}) and~(\ref{11}), are given by
\begin{equation}
\label{12}N_{\rm eq}=N_0\frac{1}{1+2 L} \quad{\rm and}\quad \sigma_{\rm eq}=N_0\frac{L}{1+2 L}.
\end{equation}

The equations of the problem, Eqs.~(\ref{9}),~(\ref{10}), and~(\ref{11}) have to be solved by taking into account the initial distribution of particles, i.e. the distribution of $N({z^*},{t^*} )$ and of its ${t^*} $-derivative at ${t^*} =0$. A simple inspection allows us to show that these two quantities are not independent. In fact, from
Eq.~(\ref{11}) it follows that
\begin{equation}
\label{c-1}\sigma({t^*} )=\frac{1}{2}\left\{N_0-\int_{-1/2}^{1/2} N({z^*},{t^*} )\,d{z^*}\right\},
\end{equation}
whose ${t^*}$-derivative is, for ${t^*} \geq 0$,

\begin{equation}
\label{c-2}\frac{d \sigma}{d{t^*} }=-\frac{1}{2}\,\,\int_{-1/2}^{1/2}\frac{\partial N}{\partial {t^*} }\,d{z^*}.
\end{equation}
In the limit ${t^*} \to 0$, Eq.~(\ref{c-2}) yields

\begin{equation}
\label{c-3}\left(\frac{d \sigma}{d{t^*} }\right)_0=-\frac{1}{2}\,\,\int_{-1/2}^{1/2}\left(\frac{\partial N}{\partial {t^*} }\right)_0\,d{z^*}.
\end{equation}
By comparing Eq.~(\ref{c-3}) with Eq.~(\ref{10-1}), one obtains

\begin{equation}
\label{c-6}\int_{-1/2}^{1/2} \left(\frac{\partial N}{\partial {t^*} }\right)_0\,d{z^*}=-2\,\,\frac{L}{A}\,\,N_s(0).
\end{equation}
As stated above, for a regular solution of our problem  $N$ and its ${t^*}$-derivative have to satisfy a condition of compatibility. In our analysis, the particles can be found in the volume or on the surface, but for ${t^*} =0$ all the particles are only in the bulk. For this reason,  we assume that for ${t^*} =0$ the initial distribution of particles  $N({z^*},0)$ is such that $N_s(0)=N(\pm 1/2,0)=0$, and $N({z^*},0)=N_0$, for ${z^*}\neq \pm 1/2$. With this assumption,  from Eq.~(\ref{c-6}) we get for the initial conditions of the problem relevant to the regular solution

\begin{equation}
\label{bc}\left(\frac{\partial N}{\partial {t^*} }\right)_0=0,\quad\quad{\rm and}\quad\quad N({z^*},0)=N_0[\theta({z^*}+1/2)-\theta({z^*}-1/2)],
\end{equation}
where $\theta(x)$ is the Heaviside's step function such that $\theta(x)=0$ for $x<0$, and $\theta(x)=1$ for $x>0$. In this framework, from Eq.~(\ref{10-1}) it follows that

\begin{equation}
\label{10-2}
\left(\frac{d \sigma}{d t^*}\right)_{t^*=0}=0,
\end{equation}
indicating that for $t^*=0$ the time derivative of the surface density of adsorbed particles is continuous.

\section{Eigenvalues and eigenvectors}
\label{Eigenvalues}
We look for a solution of Eq.~(\ref{9}), with the conditions~(\ref{10}) and~(\ref{11}), of the form
\begin{equation}
\label{13}N({z^*},{t^*} )=N_{\rm eq}+\eta({z^*},{t^*} ) \quad{\rm and}\quad \sigma({t^*} )=\sigma_{\rm eq}+s({t^*} ),
\end{equation}
where

\begin{equation}
\label{14}\lim_{{t^*} \to \infty}\eta({z^*},{t^*} )=0 \quad{\rm and}\quad \lim_{{t^*} \to\infty} s({t^*} )=0.
\end{equation}
In terms of $\eta({z^*},{t^*} )$ and $s({t^*} )$,  Eqs.~(\ref{9}),~(\ref{10}), and~(\ref{11}) read

\begin{equation}
\label{15}B\frac{\partial^2 \eta}{\partial {t^*} ^2}+\frac{\partial \eta}{\partial {t^*} }=\frac{\partial^2 \eta}{\partial {z^*}^2}\,,
\end{equation}
\begin{equation}
\label{16}A\frac{ds}{d{t^*} }=L \eta-s,
\end{equation}
\begin{equation}
\label{17}\int_{-1/2}^{1/2}\,\eta({z^*},{t^*} )\,d{z^*}+2 s({t^*} )=0.
\end{equation}
In Eq.~(\ref{15}),  $B$ is a small parameter, since we expect that the velocity of transmission of the information is finite, but large.
However, the limit operation of $B\to 0$ cannot be directly performed because in this limiting operation the differential equation passes
from the hyperbolic to the parabolic type. The point $B=0$ is thus a singular point~\cite{Malley}.
Equations~(\ref{9}),~(\ref{10}), and~(\ref{11}) can be solved by separating the variables. By putting

\begin{equation}
\label{18}\eta({z^*},{t^*} )=U({z^*}) V({t^*} ),
\end{equation}
Equation (\ref{15}) can be rewritten as

\begin{equation}
\label{19}\frac{1}{V}\left(B\,\frac{d^2 V}{d{t^*} ^2}+\frac{d V}{d{t^*} }\right)=\frac{1}{U}\,\frac{d^2 U}{d{z^*}^2}=-\alpha^2,
\end{equation}
where the separation constant $\alpha^2$ has its  real part positive, in such a manner that Eqs.~(\ref{14}) are verified. From
Eq.~(\ref{19}) we get
\begin{eqnarray}
\label{20}B\,\frac{d^2 V}{d{t^*} ^2}+\frac{d V}{d{t^*} }+\alpha^2 V&=&0\\
\label{21}\frac{d^2 U}{d{z^*}^2}+\alpha^2 U&=&0.
\end{eqnarray}
Since Eq.~(\ref{20}) has constant coefficients, its solution are of the kind $V=P\,\exp(\mu {t^*} )$, where the characteristic  exponents are
given by

\begin{equation}
\label{22}\mu_{1,2}(\alpha)=-\frac{1}{2B}\left(1\pm \sqrt{1-4 \alpha^2 B}\right).
\end{equation}
Note that, for a fixed $\alpha$, in the limit $B\to 0$, from Eq.~(\ref{22}) we get
$\mu_1\to -1/B,$  and $\mu_2\to -\alpha^2$.
This means that in the considered limit $\mu_1\to -\infty$, whereas $\mu_2$ tends to a finite quantity.

The function $V({t^*} )$ we are looking for is
\begin{equation}
\label{24}V_{\alpha}({t^*} )=P_{1\alpha}\,e^{\mu_1 {t^*} }+P_{2\alpha}\,e^{\mu_2 {t^*} },
\end{equation}
where $P_{1\alpha}$ and $P_{2\alpha}$ are integration constants to be determined.
By taking into account the symmetry of the problem, the solution of Eq.~(\ref{21}) is
\begin{equation}
\label{25}U_{\alpha}({z^*})=Q_{\alpha} \cos(\alpha {z^*}),
\end{equation}
where $Q_\alpha$ is a new constant to be determined.
The solution of Eq.~(\ref{15}) can be expressed in the form
\begin{equation}
\label{26}\eta_{\alpha}({z^*},{t^*} )=\eta_{1\alpha}({z^*},{t^*} )+\eta_{2\alpha}({z^*},{t^*} ),
\end{equation}
where
\begin{equation}
\label{27}\eta_{1\alpha}({z^*},{t^*} )=S_{1\alpha}\cos(\alpha {z^*})\,\,e^{\mu_1 {t^*} },
\end{equation}
with $S_{1}(\alpha)=P_{1,\alpha} Q_{\alpha}$, and similar relations being valid for $\eta_{2\alpha}({z^*},{t^*} )$ and $S_{2\alpha}$. The functions
$\eta_{1\alpha}$ and $\eta_{2\alpha}$ are two linearly independent solutions.

Let us consider now Eq.~(\ref{16}). After integration we get
\begin{equation}
\label{28}s({t^*} )=e^{-{t^*} /A}\,\left\{\frac{L}{A}\,\int_{0}^{t^*} \,e^{t/A}\,\eta(1/2,t)\,dt +M\right\}.
\end{equation}
This relation, written for the mode $\eta_{1\alpha}({z^*},{t^*} )$, after taking into account Eq.~(\ref{27}), becomes

\begin{eqnarray}
\label{29}
s_{1\alpha}&=&\frac{L}{1+\mu_1 A}\,S_{1\alpha}\cos(\alpha/2)\,e^{\mu_1 {t^*} }+\nonumber\\
&+&\left\{M_{1\alpha}-\frac{L}{1+\mu_1 A}\,S_{1\alpha}\cos(\alpha/2)\right\}\,e^{-{t^*} /A}.
\end{eqnarray}
A similar relation holds for $s_{2\alpha}$. By imposing now the condition (\ref{17}) for the mode $\eta_{1\alpha}({z^*},{t^*} )$ and
$\eta_{2\alpha}({z^*},{t^*} )$ we get the eigenvalues equations
\begin{eqnarray}
\label{30}f_1(\alpha') = \frac{1}{\alpha'}\,\tan(\alpha'/2)+\frac{L}{1+\mu_1 A}&=&0\\
\label{31}f_2(\alpha'') = \frac{1}{\alpha''}\,\tan(\alpha''/2)+\frac{L}{1+\mu_2 A}&=&0,
\end{eqnarray}
for the eigenvalues $\alpha'$ and $\alpha''$, and
\begin{eqnarray}
\label{32}M_{1\alpha'}&=&\frac{L}{1+\mu_1 A}\,S_{1\alpha'}\,\cos(\alpha'/2)\\
\label{33}M_{2\alpha''}&=&\frac{L}{1+\mu_2 A}\,S_{2\alpha''}\,\cos(\alpha''/2),
\end{eqnarray}
for the integration constants appearing in Eq.~(\ref{28}) for the two modes. A simple analysis allows us to verify that the eigenfunctions
$\varphi_{\alpha}=\cos(\alpha {z^*})$ are not orthogonal in $-1/2\leq {z^*} \leq 1/2$. In fact for two different $\alpha$, solutions of
Eq.~(\ref{30}), we call $\alpha_a$ and $\alpha_b$, we have
\begin{eqnarray}
\label{34} (\varphi_a,\varphi_b)&=&\int_{-1/2}^{1/2} \cos(\alpha_a {z^*})\cos(\alpha_b {z^*})\,d{z^*}=\nonumber\\
&=&-2\frac{L}{\alpha_a^2-\alpha_b^2}\left\{\,\frac{\alpha_a^2}{1+\mu_1(\alpha_a)A}-\frac{\alpha_b^2}{1+\mu_1(\alpha_b)A}\right\}
\cos\left(\frac{\alpha_a}{2}\right)\,\cos\left(\frac{\alpha_b}{2}\right)\neq 0.
\end{eqnarray}
Hence, to use the eigenfunctions we have to orthogonalize the set of $\varphi_{\alpha}=\cos (\alpha {z^*})$.

\begin{figure}
\centering
\includegraphics*[scale=.75,angle=0]{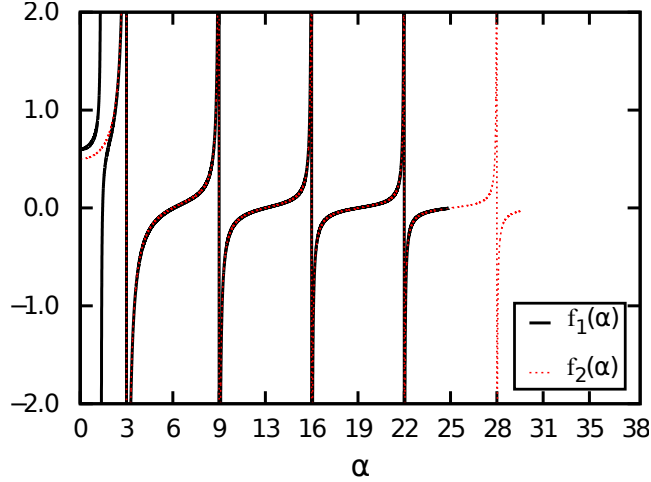}
\caption{{Behavior of $f_1(\alpha)$, given by Eq.~(\ref{30}) (black), and  of $f_2(\alpha)$, given by Eq.~(\ref{31}) (red),   versus $\alpha$ showing the zeros corresponding to the eigenvalues $\alpha'$ and $\alpha''$, for $A = 0.5$, $B= 10^{-3}$, and $L=0.1$}.}
\label{Fig_Cafasso_1}
\end{figure}

\begin{figure}
\includegraphics*[scale=.75,angle=0]{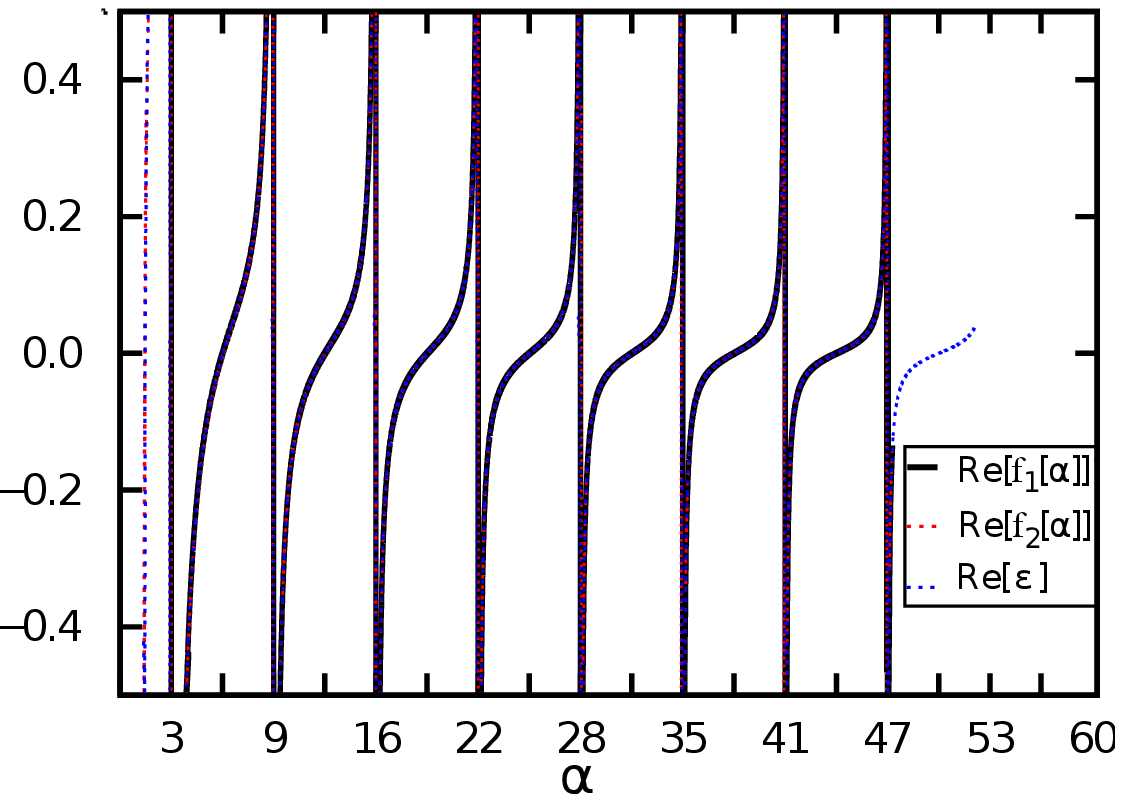}
\includegraphics*[scale=.75,angle=0]{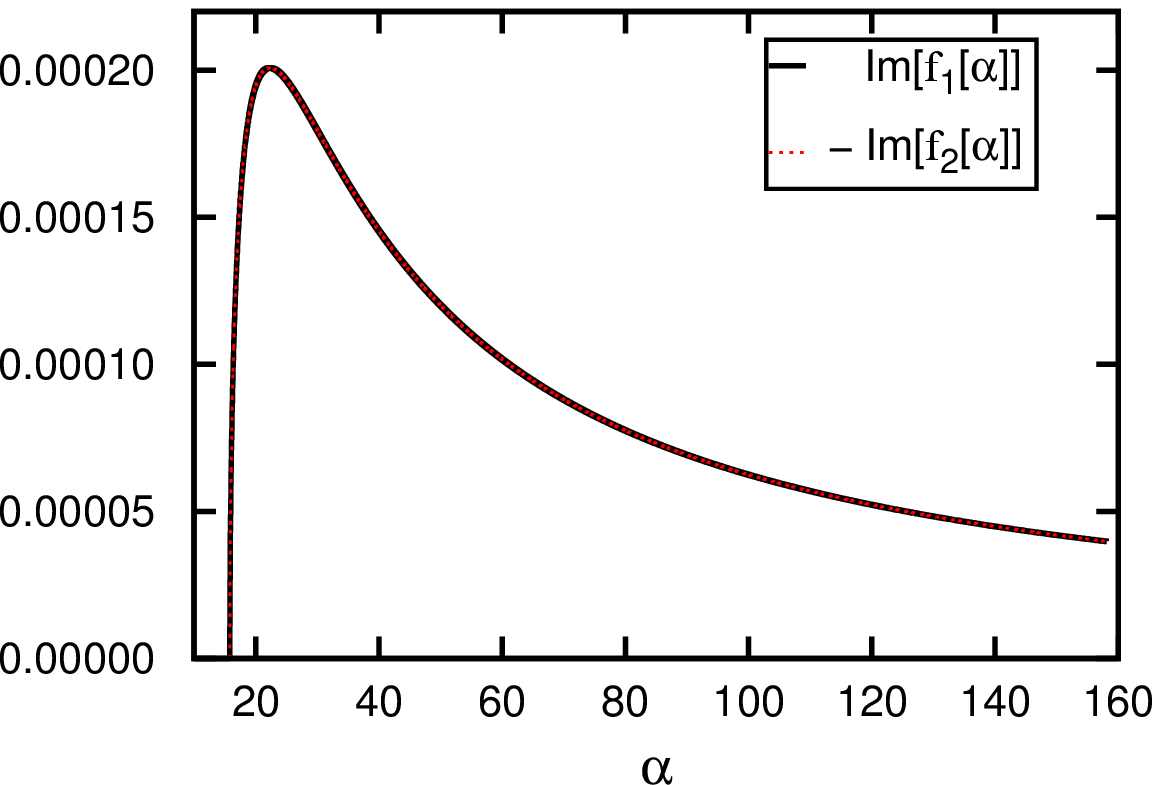}
\caption{{ a) Behavior of the real part of Eqs.~(\ref{30}) (thick, black),  (\ref{31}) (red, dashed), and ${\rm Re}[{\cal E}]$ (dotted, blue), as given by Eq.~(\ref{y-2}), versus $\alpha$  for $A =0.5$, $B= 10^{-3}$, and $L=0.1$. (b) The same for the imaginary
part of Eqs.~(\ref{30}) (thick, black) and the negative of the imaginary part of Eqs.~(\ref{31}) (dashed, red).} }
\label{Fig_Cafasso_2}
\end{figure}

By taking into account Eq.~(\ref{22}), it follows that the critical exponents $\mu_{1,2}$ are real only for $\alpha<1/(2\sqrt{B})$. Since
the physical meaning of $\alpha^2$ is a relaxation time, it follows that the dimensionless relaxation time related to the phenomenon under
investigation has a critical behavior around $4 B$, or, in absolute units, around $4 \tau_r$. Note that, for $\alpha >1/(2\sqrt{B})$,
Eqs.~(\ref{30}) and~(\ref{31}) have no solutions. This means that the number of eigenvalues is finite for $B\neq 0$. Consequently the
number of eigenfunctions is finite too, and the relevant set of eigenfunctions is not complete. It follows that is impossible to satisfy
the initial boundary conditions of the problem if one tries to solve it by means of the separation of variables. Of
course, if $B$ is small enough, and the number of eigenvalues large, an approximated solution can be found. In Fig.~\ref{Fig_Cafasso_1} we show, for a given
set of $A$, $B$, and $L$ the eigenvalues determined by means of Eqs.~(\ref{30}) and~(\ref{31}). As it is clear the eigenvalues $\alpha'$
and $\alpha''$ are close to $\alpha_m=2 m \pi$, where $m$ is an integer.
As underlined above, for $\alpha>1/(2\sqrt{B})$ the characteristics exponents are complex and conjugated, and given by

\begin{equation}
\label{y-1}\mu_{1,2}=-\frac{1}{2B}\left(1\pm i\sqrt{4 \alpha^2 B-1}\right).
\end{equation}
In this framework the real and imaginary parts of the eigenvalue equation ${\cal E}$ are

\begin{eqnarray}
\label{y-2}{\rm Re}[{\cal E}]&=&\frac{1}{\alpha}\,\tan\left(\frac{\alpha}{2}\right)+L\,\,\frac{2B(2B-A))}{(2B-A)^2+A^2(4 \alpha^2 B-1)}\\
\label{y-3}{\rm Im}[{\cal E}]&=&\pm \frac{2 B A \sqrt{4 \alpha^2 B-1}}{(2 B-a)^2+A^2(4 \alpha^2 B-1)},
\end{eqnarray}
where $+$ and $-$ refer to $\alpha'$ and $\alpha''$, respectively.
A simple inspection shows that ${\rm Re}[{\cal E}]=0$ has infinite solutions, whereas ${\rm Im}[{\cal E}]=0$ has no solutions. In Fig.~\ref{Fig_Cafasso_2} we show
${\rm Re}[{\cal E}]$ and ${\rm Im}[{\cal E}]$ versus $\alpha$. As it is clear from the quoted figure, for a given $\alpha$, different from the
solution of ${\rm Re}[{\cal E}]=0$, ${\rm Re}[{\cal E}]\gg {\rm Im}[{\cal E}]$. In addition, we underline that,
for $\alpha<1/(2\sqrt{B})$,  the eigenvalues of Eq.~(\ref{31}) are very close to ones of
${\rm Re}[{\cal E}]=0$. The same conclusion holds true for the eigenvalues of Eq.~(\ref{30}), except for the first one. For this reason in the
following we neglect the small imaginary part of the eigenvalue equation, and assume that the solutions of ${\rm Re}[{\cal E}]=0$ represent an
approximation for the eigenvalues of the problem, that we indicate simply by $\alpha$.

\section{Initial conditions}
\label{Initial}
The solution of the problem under study, due to the linear character of Eq.~(\ref{15}) and of the conditions~(\ref{16}) and~(\ref{17}), is

\begin{equation}
\label{35}\eta({z^*},{t^*} )=\sum_{\{\alpha\}}\left\{S_{1,\alpha}\,\exp[\mu_1(\alpha) {t^*} ]+S_{2,\alpha}\,\exp[\mu_2(\alpha) {t^*} ]\right\}\,\cos(\alpha
{z^*}),
\end{equation}
where $\alpha$ are the solutions of ${\rm Re}[{\cal E}]=0$, $\mu_{1,2}(\alpha)$ given by Eqs.~(\ref{22}), and $S_{1,\alpha_1}=0$ for the reason
discussed above. The coefficients $S_{1,\alpha}$ and $S_{2,\alpha}$ have to be determined by means of the initial conditions on $N({z^*},{t^*} )$
and $\partial N/\partial {t^*} $ for ${t^*} =0$.

\begin{figure}
\centering
\includegraphics*[scale=.75,angle=0]{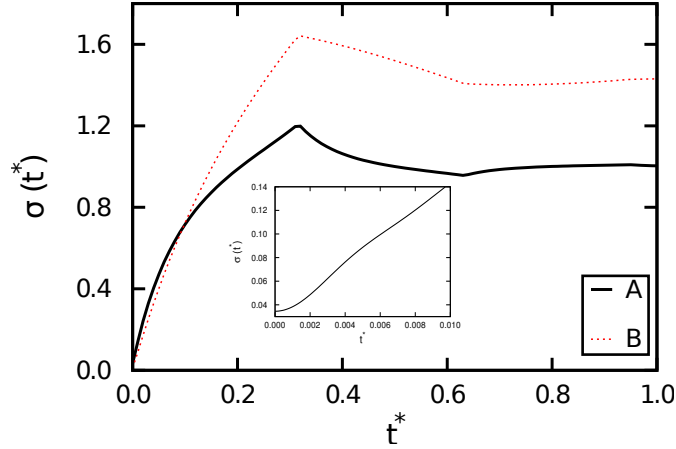}
\caption{ $\sigma(t^*)$ vs. $t^*$  for $B = 0.1$ and $A = 10^{-3}$ for two illustrative situations: $L=1.0$ (black) and $L=10$ (red).
The inset ($L=1.0$) shows that for $t^*=0$ the time derivative of the surface density of adsorbed particles is continuous, as predicted by Eq.~(\ref{10-2}). The summations in Eq.~(\ref{sigmat*}) have been evaluated for the first fifty eigenvalues.}
\label{Fig_Cafasso_3}
\end{figure}

The initial conditions on $\eta({z^*},{t^*} )$ are such that
\begin{equation}
\label{37}\eta({z^*},0)=N({z^*},0)-N_{\rm eq}\quad{\rm and}\quad \left(\frac{\partial \eta}{\partial {t^*} }\right)_{{t^*} =0}=0,
\end{equation}
where $N({z^*},0)$ is defined in Eq.(\ref{bc}). By means of  Eq.~(\ref{35}),  Eq.(\ref{37}) can be rewritten as

\begin{eqnarray}
\label{38}\sum_{\{\alpha\}}\left(S_{1,\alpha}+S_{2,\alpha}\right)\,\cos(\alpha {z^*})&=&N({z^*},0)-N_{\rm eq}\\
\label{39}\sum_{\{\alpha\}}\left[\mu_1(\alpha) S_{1,\alpha}+\mu_2(\alpha) S_{2,\alpha}\right]\,\cos(\alpha {z^*})&=&0.
\end{eqnarray}
From Eq.~(\ref{39}) it follows that

\begin{equation}
\label{40}S_{2,\alpha}=-\frac{\mu_1(\alpha)}{\mu_2(\alpha)}\,S_{1,\alpha},
\end{equation}
and Eq.~(\ref{38}) becomes

\begin{equation}
\label{41}\sum_{\{\alpha\}}S_{1,\alpha}\left[1-\frac{\mu_1(\alpha)}{\mu_2(\alpha)}\right]\,\cos(\alpha {z^*})= \sum_{\{\alpha\}} C_{\alpha}
\varphi_{\alpha}({z^*})= N({z^*},0)-N_{\rm eq}.
\end{equation}
Equation~(\ref{41}) has to be inverted to determine $S_{1\alpha}$, and then  $S_{2\alpha}$, by means of which we can evaluate $\eta({z^*},{t^*} )$
and, finally, $s({t^*} )$. As stated above,  this is a difficult task since the eigenfunctions $\varphi_{\alpha}({z^*})$ are not orthogonal. To
orthogonalize them, we assume that it is possible to expand $\varphi_{\alpha}({z^*})$ in terms of an orthogonal set $\psi_{\alpha}({z^*})$ such
that

\begin{equation}
\label{U}
\varphi_{\alpha}({z^*}) = \sum_{\alpha} U_{\alpha l} \psi_l({z^*}).
\end{equation}
In this manner, Eq.~(\ref{41}) can be rewritten as

\begin{equation}
\label{eqr} R_j = \sum_{\alpha} U_{\alpha j}\, C_{\alpha},
\end{equation}
where

$$
R_j = \frac{\int_{-1/2}^{1/2} [N({z^*},0)-N_{\rm eq}] \psi_j({z^*}) d{z^*}}{\int_{-1/2}^{1/2} \psi_j({z^*}) \psi_j({z^*}) d{z^*}}.
$$
 In matrix notation, Eq.~(\ref{eqr}) becomes
 ${\bf R} = {\bf U}^T\, {\bf C}$,  from which it follows that
${\bf C} = {\bf V}^T\, {\bf R}$, where ${\bf V} = {\bf U}^{-1}$. To
obtain the elements of $\bf V$, we can implement a way that is more
suitable to be numerically handled~\cite{Morse}, namely

\begin{equation}
\psi_q({z^*}) = \sum_{\alpha=1}^q \frac{M_{\alpha q}}{M_{qq}}\,
\varphi_{\alpha}({z^*})= \sum_{\alpha=1}^q V_{\alpha q} \varphi_{\alpha}({z^*}),
\end{equation}
where $M_{\alpha q}$ is the minor of the element

$$d_{\alpha q} = \int_{-1/2}^{1/2} \varphi_{\alpha}({z^*}) \varphi_q({z^*})\,d{z^*}$$
 in the determinant $D_{q}$ defined as

\begin{eqnarray*}
D_1 &=& d_{11}\nonumber \\
D_2 &=& \left |\matrix{ d_{11} & d_{12} \cr d_{21} & d_{22} \cr  } \right| \nonumber \\
D_3 &=& \left|\matrix{ d_{11} & d_{12} & d_{13} \cr d_{21} & d_{22}
& d_{23} \cr d_{31} & d_{32} & d_{33} \cr  }\right|; \quad {\rm
etc.}
\end{eqnarray*}
This procedure allows us to obtain the coefficients $C_{\alpha}$ and, then, using~(\ref{40}), $S_{1, \alpha}$ and $S_{2, \alpha}$ giving
the solutions $\eta({z^*},{t^*} )$ and $s({t^*} )$ in closed analytical form.  In particular, the surface density of particles may be rewritten as:

\begin{equation}
\label{sigmat*}
\sigma(t^*) = \sigma_{\rm eq} - \sum_{\alpha} \frac{1}{\alpha} \left[ S_{1\alpha} e^{\mu_1(\alpha)t^*} + S_{2\alpha} e^{\mu_2(\alpha)t^*}\right]\, \sin(\alpha/2),
\end{equation}
and is shown in Fig.~\ref{Fig_Cafasso_3} for two illustrative cases.  These results
are impressive. In the framework of Langmuir's approximation for the adsorption phenomenon the expected temporal behavior of $\sigma(t)$ is monotonous, with the density reaching a saturation value for large enough time in the parabolic approximation for the diffusion equation. The presence of the second derivative in Eq.~(\ref{6}) (accounted for by $B\ne0$) is clearly responsible for the oscillating behavior of $\sigma$ shown in Fig.~\ref{Fig_Cafasso_3}. As mentioned before, this results is in good qualitative agreement with the experimental data reported in~\cite{Cosgrove}. The
non-monotonic behavior found here is strongly dependent on the values of the parameters $A$ and $L$, but the choice of the value of $B$ is crucial in determining this behavior, as it will be discussed
in details in Sec.~\ref{Results}. The slope of $\sigma(t^*)$ at the origin is practically independent of the value of $L$. Likewise, the position of the maxima of $\sigma(t^*)$ are also essentially independent of $L$. Notice, however, that the value of $\sigma(t^*)$ at the maximum is clearly sensible to the value of $L$, as will be discussed below.

\section{Numerical procedure}
\label{Numerical}

The finite difference based procedure exploited to solve the problem described by Eqs.~(\ref{9}),~(\ref{10}),  and~(\ref{11}) is now outlined. Thanks to the symmetry, only half geometry can be considered. Let us thus partition the spatial domain $0\leq {z^*} \leq 0.5$ into $n_{z^*}$ segments of length $h=0.5/n_{z^*}$, and the time domain $[0,T]$ under consideration into $n_{t^*} $ segments of length $k=T/n_{t^*} $. By assuming that $N({z^*}_i,{t^*}_j)=N_{i,j}$ and that $\lambda=k/h$, the numerical algorithm for the inner points of the spatial domain ($i=2,...,n_{z^*}$) reads:
\begin{itemize}

\item  $j=0$: $$N_{i,0}=N_0$$

\item $j=1$ ($g_i=(\frac{\partial N}{\partial t})_{i,0}$):

$$N_{i,1}=\frac{\lambda^2}{2B}(N_{i+1,0}+N_{i-1,0})+\frac{B-\lambda^2}{B}N_{i,0}+\frac{ (2B-k)k}{2B}g_{i}$$

\item  $j=2,...,n_{t^*} +1$

$$ N_{i,j}=\frac{2 \lambda^2}{2B+k}(N_{i+1,j-1}+N_{i-1,j-1})+\frac{4 (B-\lambda^2)}{2B+k}N_{i,j-1}-\frac{2B-k}{2B+k} N_{i,j-2}.$$

\end{itemize}

For what concerns the boundary conditions ($i=1$ and $i=n_{z^*}+1$), the following relationship is imposed to satisfy the symmetry condition:

$$N_{1,j}=N_{2,j}.$$
On the other hand, as regards $N_{n_{z^*}+1,j}$, its values are obtained at each time $j$ by inserting Eq. (\ref{11}) into (\ref{10}) and approximating the integrals by means of the simple trapezoidal rule:

$$\int_{a}^{b} f(x)\, dx \approx \frac{h}{2} \sum_{k=1}^{n_{z^*}} \left( f(x_{k+1}) + f(x_{k}) \right).$$

In the procedure described above it is important to underline that the numbers of spatial and
time intervals, $n_{z^*}$ and $n_{t^*} $, respectively, should be increased till
convergence occurs. Moreover, $\lambda$ must be kept sufficiently small
in order to prevent numerical instabilities. The problem is more complicated than that related to
classical differential equations, e.g. the wave equation, according to which the condition $\lambda<1$ must be satisfied.
This is due to the nonlocal condition \ref{11} coupled with \ref{10}: at each time, the density $N(0.5,t^*)$ depends on the solution
throughout the sample. Indeed, for what concerns the results presented in Sec.~\ref{Results}, it is found that the value of $\lambda$
is affected by the parameters considered in the analysis, and especially by $B$: the lower is $B$, the lower must be $\lambda$.
For $B=10^{-3}$, $\lambda$ is be assumed to be equal to $2.5 \times 10^{-2}$.

\section{Results}
\label{Results}

The numerical procedure discussed in Sec.~\ref{Numerical} is now implemented for a more detailed investigation of the behavior of $\sigma(t^*)$ and $N({z^*}, t^*)$ as a function of the characteristic times $\tau_r$, $\tau_D$, $\tau_a$ and lengths $d$ and $\kappa_a \tau_a$. By comparing the results with those presented in Fig.~\ref{Fig_Cafasso_3}, the approximated method to obtain the eigenvalues of the problem by solving Re$[{\cal E}]=0$ presented in Sec.\ref{Eigenvalues} is validated.

\begin{figure}
\centering
\includegraphics*[scale=.65,angle=0]{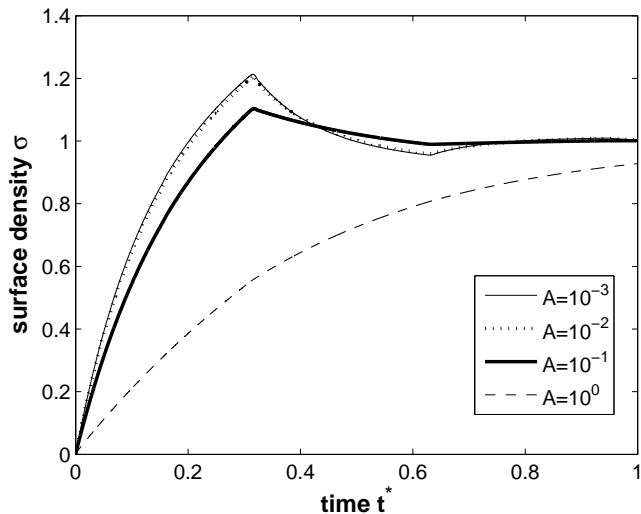}
\caption{$\sigma(t^*)$ vs. $t^*$  for $N_0$, $B = 0.1$ and $L= 1$ for different values of $A$.}
\label{Fig_Cafasso_4}
\end{figure}
\begin{figure}
\centering
\includegraphics*[scale=.65,angle=0]{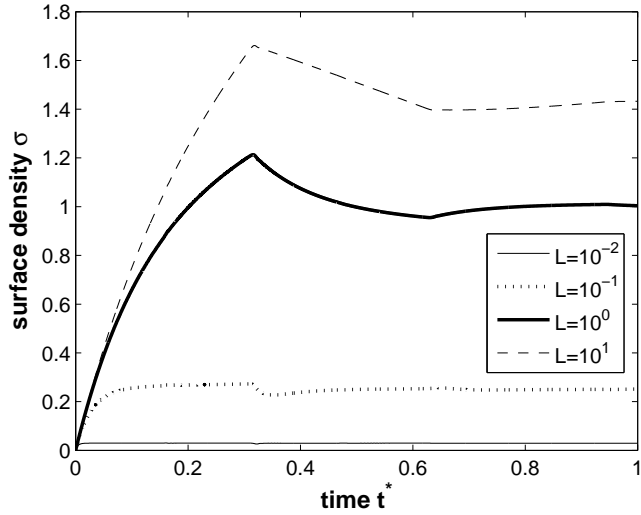}
\caption{$\sigma(t^*)$ vs. $t^*$  for $N_0 =3$,  $B = 0.1$, and $A=0.01$ for different values of $L$.}
\label{Fig_Cafasso_5}
\end{figure}
\begin{figure}
\centering
\includegraphics*[scale=.65,angle=0]{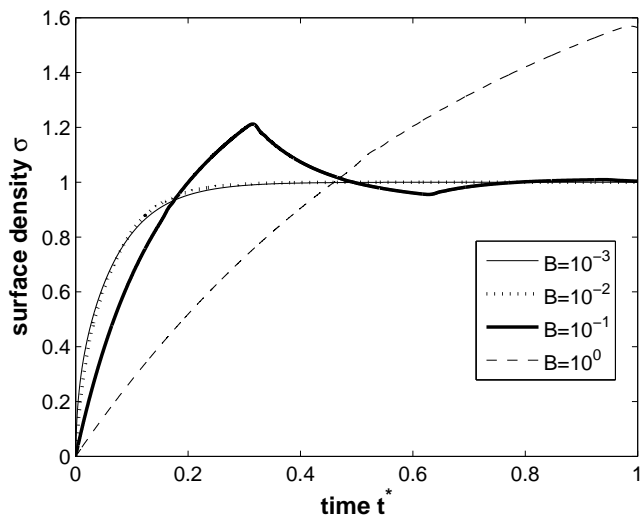}
\caption{$\sigma(t^*)$ vs. $t^*$  for $N_0 =3$, $A=0.01$, and $L=1.0$ for different values of $B$.}
\label{Fig_Cafasso_6}
\end{figure}

 In Fig.~\ref{Fig_Cafasso_4}, the surface density of particles is shown as a function of
$t^*$ for different ratios between the characteristic desorption time $\tau_a$ and the diffusion time $\tau_D$.
For $\tau_a < \tau_D$ ($A < 1.0$), the desorption process is significant for initial times and some kind of ``competing effect'' with adsorption and diffusion may be found at the surface. This competition for short times is probably the main mechanism underlying the nonmonotonic trend of $\sigma$. In Fig.~\ref{Fig_Cafasso_5}, the varying quantity is the ratio between a characteristic ``adsorption length'' (represented by $\ell = \kappa_a \tau_a$) and the thickness of the sample.
For $\ell \ll d $ ($L \ll 1)$, the adsorption-desorption process occurs over a very short distance, i.e., it is strongly localized near the surface and is not conspicuous. The number of particles on the surface is very small. As $L$ increases, also the ``adsorption length'' increases and an increasingly greater number of particles takes part into the adsorption-desorption phenomena. For $L \gg 1$, it is expected that the adsorption-desorption phenomena involve all the particles in the sample. This explains the high value of the surface density and also
the different values of $\sigma$ at the maxima found in
Fig.~\ref{Fig_Cafasso_3} when $L\gg 1$. In all the cases in which the desorption process is present, the non-monotonic behavior is assured by the small value of $B$. Indeed, in Fig.~\ref{Fig_Cafasso_6}, the values of $B$ are
chosen to illustrate the role of the second derivative in Eq.~(\ref{6}). For very small values of $B$ ($B=10^{-3}$ and $B= 10^{-2}$) the adsorption phenomenon presents the expected monotonic behavior of Langmuir's approximation. As $B$ increases, the oscillating behavior arises in the system and is very clear for the initial times when $B = 0.1$ but is also present if one waits more time, i.e., when $B=1.0$ the maxima in the density are found for
$t \gg \tau_D$. In this later case, the characteristic time $\tau_r = \tau_D$,  which implies  that the velocity $c$ of the density wave becomes small. However,  since the whole sample takes part into the adsorption-desorption phenomenon (because $\ell = d$) the surface density increases before starting to oscillate for large $t$.

Another feature of the behavior of the
surface density can be quantitatively understood from the previous figures. The velocity of the density wave, in the units we are using here, is given by $c = 1/\sqrt{B} \approx 3.16$, if $B=0.1$. Now, if we consider the curve $\sigma(t^*)$ vs. $t^*$, the first maximum may be found for $t^* = T \approx 0.32$, i.e, $1/c$.  This is an expected result: when the concentration starts to vary in view of the adsorption phenomenon on the surface ($z^*=-1/2$, for instance), the more distant particles are located close to the second surface ($z^*=1/2$), and they have to cover the distance 1  with velocity $c$.
For this reason, the next maximum will be found after a time interval $T$. Finally, one notices that the slope of $\sigma(t^*)$ at the origin is of the order of  $N_0 c=N_0/\sqrt{B}$. This result can be easily understood by taking into account that the bulk density of particles due to the drift related to the presence of the adsorbing surface is $j=N c$. For $t^*=0$ the bulk density of particles just in front to the surface is $N_0$, and hence $j(0)=N_0 c$. Since this current density is responsible for the increasing of the surface density of particles, in a first approximation, by neglecting the diffusion phenomenon, $j=d \sigma/d t^*$. Consequently the initial time derivative of the surface density of adsorbed particles is $(d \sigma/d t^*)_0=N_0 c=N_0/\sqrt{B}$, in agreement with the results reported in Figs.~\ref{Fig_Cafasso_3},~\ref{Fig_Cafasso_4}, 
and~\ref{Fig_Cafasso_5}, corresponding to the same value of $B$. On the contrary from Fig.~\ref{Fig_Cafasso_6}  it is possible to verify that changing $B$, $(d \sigma/d t^*)_0\propto 1/\sqrt{B}$.

\begin{figure}
\centering
\includegraphics*[scale=.65,angle=0]{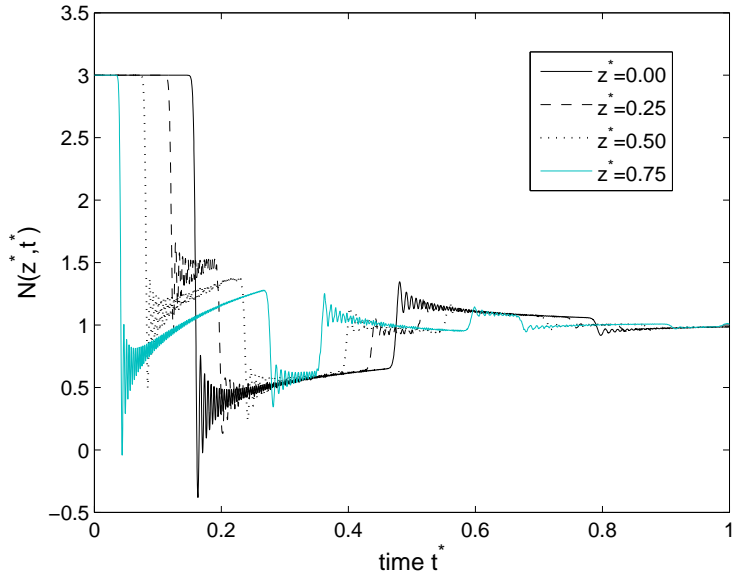}
\caption{$N({z^*}, t^*)$ vs. $t^*$  for $N_0 =3$, $A=0.01$, $B = 0.1$, and $L=1.0$ at different positions inside the sample. }
\label{Fig_Cafasso_7}
\end{figure}
\begin{figure}
\centering
\includegraphics*[scale=.65,angle=0]{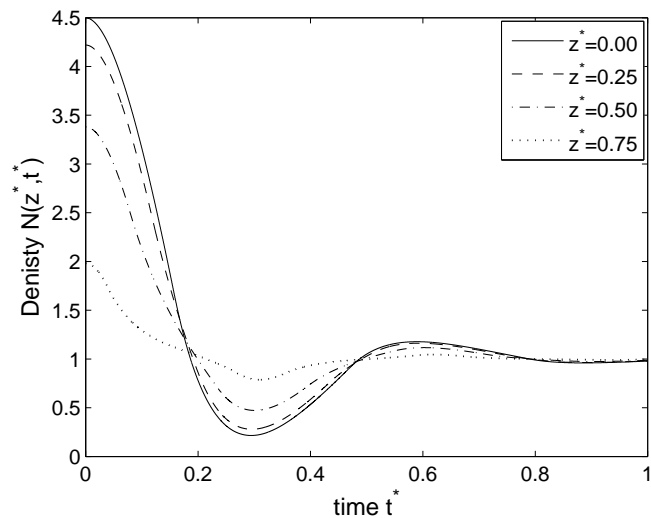}
\caption{The same as in Fig.~\ref{Fig_Cafasso_7} for a parabolic (smooth) initial distribution. }
\label{Fig_Cafasso_8}
\end{figure}

 The behavior of the bulk density of particles is shown in Fig.~\ref{Fig_Cafasso_7} in different positions inside the sample. The initial condition is such that $N({z^*}, 0) = N_0$ for
$|{z^*}| \ne 1/2$.  The same considerations done on the behavior of $\sigma$ can be useful here to interpret the behavior of the bulk density. For instance, if we consider the position ${z^*}=0$, the distance from the surface is $1/2$. Thus, the bulk density changes after an interval of $ T^*=0.5/c \approx 0.16$, as can be easily checked on Fig.~\ref{Fig_Cafasso_7}. In the same manner, having in mind the positions ${z^*}=0.25$ and ${z^*}=0.75$, one notices that the maximum in the density may be found after a time interval $1/c$,  because the density waves have to go towards the surface and to come back. If we consider, for simplicity, the position ${z^*}=0$, we notice that the density changes after a time interval $T^*$, when the particles start to move towards the surface. After a time $T^*$,  the particles reach the surface at which part of them is reflected (the other part may be adsorbed). This reflected part arrives again at the position ${z^*}=0$, after a time $T^*$,  where they interfere with the particles coming from the opposite surface. Consequently, at the time $T^*$ the density starts to change and after a time $2T^*$ it is recomposed, and so on. For very large time intervals, however, the surface density tends to a saturation while the bulk density tends to an almost constant value. The oscillations of the bulk density versus $t^*$, evident in Fig.~\ref{Fig_Cafasso_7}, are related to numerical problems in facing the
discontinuous initial conditions (\ref{bc}). For smoother distributions
$N(z^*,0)$, this oscillating behavior is no longer present: in 
Fig.~\ref{Fig_Cafasso_8} we
show the evolution of the bulk density related to a parabolic initial
distribution.

\section{
Conclusions}

The diffusion of particles in a finite-length sample is described here by a diffusion equation of hyperbolic type (Cattaneo's equation). The solution of this equation is subjected to boundary conditions involving a kinetic balance equation at the surfaces.  The problem was analytically solved by means of separation of variables, invoking a detailed process for the orthogonalization of the eigenfunctions of the problem. In addition, a detailed numerical analysis allowed the exploration of the role of the parameters of the model on the temporal behavior of the  bulk and surface density of particles. In contrast with the temporal behavior usually found in solving  the diffusion equation of  parabolic type (usual Fick's law), the solutions show a remarkable oscillatory behavior,  both in bulk and in the surface, for the initial times. This kind of formalism may be  helpful to explore memory effects on the adsorption-desorption phenomenon at the limiting surfaces as well as on the bulk diffusion of neutral particles.

\acknowledgments

Many thanks are due to L. Pandolfi and A. Scarfone for useful discussions.
The research leading to these results has received funding from the European Research Council under the European Union's Seventh Framework Programme (FP/2007-2013)/ERC Grant Agreement No. 306622 (ERC Starting Grant Multi-field and multi-scale Computational Approach to Design and Durability of PhotoVoltaic Modules -CA2PVM). The support of the Italian Ministry of Education, University and Research to the Project FIRB 2010 Future in Research Structural mechanics models for renewable energy applications (RBFR107AKG) is gratefully acknowledged.

\end{document}